\documentclass[12pt,preprint]{aastex}
\usepackage{bbm}
\usepackage{natbib}

\begin{document}

\title{Signatures of Cosmic Strings in the Cosmic Microwave Background}

\author{A. S. Lo \& E. L. Wright}
\affil{UCLA Astronomy, PO Box 951562, Los Angeles, CA. 90095-1562 U.S.A.}
\email{amy@astro.ucla.edu, wright@astro.ucla.edu}
\begin{abstract}
We report a search for signatures of cosmic strings in the the Cosmic
Microwave Background data from the Wilkinson Microwave Anisotropy
Probe.  We used a digital filter designed to search for individual
cosmic strings and found no evidence for them in the WMAP CMB
anisotropies to a level of $\Delta T/T \sim 0.29$ mK.  This
corresponds to an absence of cosmic strings with $ G\mu \ga 1.07
\times 10^{-5}$ for strings moving with velocity $v = c/\sqrt{2}$.
Unlike previous work, this limit does not depend on an assumed string
abundance.  We have searched the WMAP data for evidence of a cosmic
string recently reported as the CSL-1 object, and found an ``edge''
with 2$\sigma$ significance.  However, if this edge is real and
produced by a cosmic string, it would have to move at velocity $\ga$
0.94c.  We also present preliminary limits on the CMB data that will
be returned by the PLANCK satellite for comparison.  With the
available information on the PLANCK satellite, we calculated that it
would be twice as sensitive to cosmic strings as WMAP.
\end{abstract}

\keywords{(cosmology:) cosmic microwave background}

\section{Introduction}
Topological defects have been proposed as large scale structure (LSS)
candidates but have fallen out of favor due to the lack of evidence
for their existence.  Topological defects form as results of phase
transitions, many of which occurred in the first year after the big
bang.  They were ideal LSS candidates for three reasons.  First,
defects reflect the energy scale of the phase transition, therefore,
the earlier the phase transition, the higher the energy density of the
topological defect, and the larger the gravitational potential well
for structure formation.  Second, these high energy phase transitions
occurred a long time before observational evidence of the first stars
and galaxies.  This would give gravity enough time to coalesce
material to form structures.  Last, cosmic strings, the most popular
class of topological defects, are geometrically similar to the
filamentary LSS first observed in deep redshift surveys (see e.g.
\citealt{vac86}).

However, despite these attractive features, results of the COsmic
Background Explorer (COBE), an all-sky Cosmic Microwave Background (CMB)
satellite, showed that cosmic strings were not responsible for large
scale structure formation, because the CMB anisotropies were
consistent with a Gaussian signature, while numerical simulations of
all topological defects show that they would leave distinct
non-Gaussian signatures in the CMB (see, e.g., a review by
\citealt{all97}).  The COBE results were interpreted to mean that the
observed LSS's are products of perturbations with Gaussian seeds, and
cosmic strings fell out of favor as a structure formation mechanisms.

To date, there does not exist any conclusive observational evidence of
the existence of cosmic strings or any other topological defect such
as magnetic monopoles.  On the other hand, CMB data do not preclude
their existence in small enough numbers such that they do not
appreciably affect the CMB angular power spectrum. There could be on
the order of a few cosmic strings in the sky and they would not
contribute enough power to significantly alter the CMB angular power
spectrum (for recent cosmic string searches, see e.g. \citealt{per98}). 
Furthermore, topological defects are necessary products of
certain types of phase transitions in the early universe, and are
useful as possible explanations for variety of phenomenon such as
Gamma Ray Bursts (e.g. \citealt{ber01}).  This persistent interest in
topological defects motivated our search for rare cosmic strings.

If cosmic strings do exist, they need to be either few in number, or
have a small linear mass density.  With that
in mind, the current project performed a search for individual cosmic
strings based on an all-sky CMB survey data.  The data returned from
the Wilkinson Microwave Anisotropy Probe (WMAP) have an angular
resolution of 13 arc-minutes which enabled searches for individual
cosmic strings, as 2 degrees is the scale at which we expect to see
cosmic strings (see section II).  While maps with greater resolution
have been produced by balloon-borne experiments
(e.g. \citealt{bou02}), these experiments only observe a very small
area of the sky.  WMAP covers the entire $4\pi$ steradian of sky and
is therefore uniquely suited for searches of rare cosmic
strings.  Our search involves a pixel by pixel filtering, as opposed
to the wavelet analysis that some others have performed on the small
scale CMB experiments (e.g. \citealt{bar01}, \citealt{sta04}), although the underlying
principles of searching for non-Gaussian signatures are similar.

\section{Topological Defects}

We are particularly interested in using cosmic strings as
discriminators of the validity of Grand Unification Theories
(GUT). GUTs have their roots in the successes of the electroweak
union.  The Glashow-Weinberg-Salam theory \citep{wei67} showed that
electromagnetism and the weak force are united at energies slightly
larger than the masses of their force mediators. The mass of the $W$
boson is 81.4 GeV, the mass of the $Z$ boson is 91.1 GeV and the
photon is massless \citep{hag02}.  The current Standard Model of
particle physics describes the universe well up to the electroweak
unification, which occurred $\sim10^{-10}$ seconds after the Big
Bang. GUTs seek to extend the Standard Model to include unification
with the strong force, and describe all forces (except gravity) as the
result of one unifying theory.  Under these theories, the universe
went through a series of phase transitions at various critical
energies as it cooled.

At each epoch where a phase transition happens there is the
possibility of forming topological defects.  The high energies
involved in GUT phase transitions cannot currently be reproduced in
experimentally, which resulted in a lack of theoretical constrains for
GUTs.  This fact partially accounts for the proliferation of these
theories, very few of which produce testable results.  Topological
defects, however, are necessary products of certain types of phase
transitions.  Different phase transitions with different properties
produce distinct topological defects.  Even if one does not believe
that a GUT can be constructed, the existence or absence of topological
defects can still give us limits on the types of allowable phase
transitions in the early universe, and therefore more information on
the evolution of the fundamental forces at very high energies.

Cosmic strings are a class of topological defects which are
1-dimensional; they are created as either filaments which span the
horizon, or as closed loops.  Filamentous cosmic strings leave
tell-tale gravitational signatures.  The metric of a long string in
cylindrical coordinates, to linear order in mass per unit length of
the string, $\mu$, is given by,
\begin{equation}
ds^2 = dt^2 - dz^2 - (1-h)(dr^2 + r^2 d\theta^2)
\end{equation}
where $h = 8 \pi G\mu \ln (r/r_0)$, $r_0$ is a constant of
integration, and $r = (x^2 + y^2)$, where $x$ and $y$ are the
Cartesian coordinates for a string lying along the $z$-axis
\citep{vil94}.  In general, $\mu$ is expressed as a dimensionless
quantity, $G\mu$, and,
\begin{equation}
G\mu \sim \frac{\eta^2}{m^2_{pl}},
\end{equation}
where $m_{pl}$ is the Planck mass, and $\eta$ is the symmetry breaking
energy scale at which the cosmic string was formed.  This metric
describes a conical space, which can be thought of as a Euclidean
space with a wedge removed.  The angle of the removed wedge is given
by,
\begin{equation}
\Delta = 8\pi G \mu.
\end{equation}
Two photons traveling on either side of the wedge would seem to be
bent towards each other, and so cosmic strings can act as
gravitational lenses. 

The electroweak symmetry breaking scale occurs at 100 GeV. An
electroweak string would have $G\mu \sim 10^{-34}$; a GUT scale string
would have $\eta \sim 10^{16}\; \mbox{GeV}$, which gives $G\mu \sim 10^{-6}$.
Therefore, knowing the masses of the strings can give us the energy
scale of the phase transition that created them. Alternatively, the
absence of strings would give an upper limit to the energy of phase
transitions which must create cosmic strings.  Numerical simulations
have shown that strings could intersect and break off loops in a
process called intercommutation; this way, strings do not dominate the
energy density of the universe. The amount of intercommutation can be
tuned so that infinite strings have many intersections.  Closed loops
can wiggle and emit gravitational radiation, which causes the loop
radii to decrease.  Eventually, within the lifetime of the universe,
string loop radii can reach zero.  This is their primary energy
dissipation mechanism (see e.g. recent numerical simulations,
\citealt{moo02}) that prevents cosmic strings from dominating the
energy-density of the universe.  Therefore, even if we do not directly
observe a cosmic string today, their formation is not ruled out.

Cosmic strings grow along with the horizon size, and because they are
created with relativistic velocity, string wakes will span the
horizon.  At the epoch of last scattering (LS), the horizon has
angular size $\theta_H$ given by,
\begin{equation}
\theta_H \backsimeq z^{-1/2}_{LS} \sim 2 \degr.
\end{equation}
These string wakes create disturbances in the matter contained within
the horizon.  Since the CMB freezes gravitational signatures at LS, we
expect to find most of the signatures of cosmic strings in the CMB at
the LS scale.  However, different numerical simulations give different
string densities and velocities and there is little agreement on the
size and number of strings we should see in the sky.  The one fact
that all simulations agree on is that string wakes leave edge-like
signatures in the CMB.

We emphasize that cosmic strings are necessary products of certain
phase transitions.  The detection of a string would certainly be very
interesting, but the absence of these defects is equally telling. If
we expect GUT scale strings to form, one defect would be created per
causal horizon. This leads to a possibility of $\sim 13000$ cosmic
strings in CMB. The gravitational signatures of these strings will
persist in the CMB even if the strings themselves have dissipated.

\section{WMAP}

Photons produced in the early universe, when the ambient temperature
was high enough to keep hydrogen ionized, remained in equilibrium with
the baryons and traced baryon density.  At a red-shift of $z \sim
1100$, electrons and protons recombined and the universe became
transparent to photons.  These photons have redshifted as the universe
expanded and we now observe them as the 3K Cosmic Microwave
Background.  The photons retained the gravitational signatures of the
early universe density distribution, except when modulated by the
Sunyaev-Zeldovich effect.

The primary effect of a cosmic string on the CMB is to create
temperature anisotropies due to the wake of the moving string. Along
a particular line of sight, the temperature anisotropy, $\delta T/T$
induced across a string moving in the plane of the sky is given by,
\begin{equation}
\label{string_t}
\frac{\delta T}{T} = 8\pi G \mu \beta \gamma.
\end{equation}
where $\gamma$ is the Lorentz factor and $\beta = v/c$ is the velocity
of the string.

The Wilkinson Microwave Anisotropy Probe (WMAP) is the latest NASA
satellite to measure the temperature anisotropies of the CMB (see the LAMBDA
website for a complete list of publications and public release data
products\footnote{\url{http://lambda.gsfc.nasa.gov/product/map}}).
WMAP has two back-to-back Gregorian telescopes which observe two
patches of the sky separated by 141$\degr$. A set of differencing
assemblies obtains the temperature difference between the two patches.
The final data product of WMAP is the CMB temperature anisotropy of
the each pixel in the sky, except where there are microwave sources
and the pixels are masked.  These masked pixels include most of the
Galactic plane, the Galactic bulge, and some scattered sources off the
Galactic plane. 

WMAP was launched on June 30, 2001, and arrived at its L2 orbit on
Oct. 1, 2001.  The first year data release in February, 2003, contained
data taken by WMAP from Aug. 10, 2001 to Aug. 9th, 2002.  WMAP is still
taking data and is expected to last at least until 2005. The following
work is based on the first year data.  As more data are released in
the future, the sensitivity of this project will be improved as the
pixel noise is reduced by repeated observations.

WMAP observes the sky at 5 frequencies, from 23 to 94 GHz. Relevant
WMAP characteristics are tabulated in Table \ref{WMAP_char}.  The data
products released by the WMAP science team include the thermodynamic
temperatures of each WMAP pixel at all 5 frequencies.  We used the
data from the 3 highest frequency bands (Q, V, W) in our analysis.

\section{String Search: The Edgefinder}
 
For our string search, we used the WMAP map of temperature
differences, $\Delta T/T$, of the full sky.  At 13 arc-minute
resolution, the WMAP sky is divided into $12 \times 4^{9}$ pixels.
WMAP uses the Hierarchical Equal Area isoLatitude Pixelisation of the
sphere (HEALPix\footnote{\url{http://www.eso.org/science/healpix/}},
\citealt{gor99}) to pixelize the sky.  We have designed a digital
filter for the WMAP data in search of signatures of filamentous string
wakes, and we named this filter the Edgefinder.  The Edgefinder took
an input pixel and defined a window with radius $RAD$.  Each pixel
within this circular window was assigned an $[x,y]$ pair, with the
input pixel at [0,0], and the rest of the pixels' $x$s and $y$s are
the normalized (to $RAD$) displacements from the input pixel.  The
y-axis was defined to be parallel to the line connecting the North and
South Galactic poles.  Our filter window was small enough that we
could consider the sky inside to be flat.  For each pixel within the
window, the Edgefinder multiplied the filter value, $F(x,y)$, with the
pixel temperature at $[x,y]$ and stored the sum of this product for
all pixels in the window at the position of the input
pixel. Therefore, the output of this digital filter was a map where
the value at each pixel was a sum of the effects of the filter on the
surrounding pixels.  We called this output the Edgefinder value map,
EV for short.

For the WMAP data and associated simulations, the Edgefinder had an
$RAD = 1 \degr$ window in the sky, which corresponded to the size of
the horizon at LS.  This window had a radius of 18 pixels and usually
contained ~260 pixels in total.  The filter value, $F(x,y)$, was
designed to pick out edges in the sky aligned with the y-axis of the
filter.  The specific values of each of the filter was generated
separately, because pixel centers were slightly offset from each other
depending on the pixel latitude, and the edges of the filter were
ragged due to the diamond grid of the HEALPix scheme. This, coupled
with the fact that we only have 240 pixels in the window, meant that
the sum of the filter values didn't always sum to a perfect zero.  We
therefore generated the filter values twice at each pixel: the first
time to collect the excess, $a$, for each window, and the second time
to distribute $a$ evenly among the pixels. This forced every window to
sum to zero exactly.  

\begin{figure}
\plotone{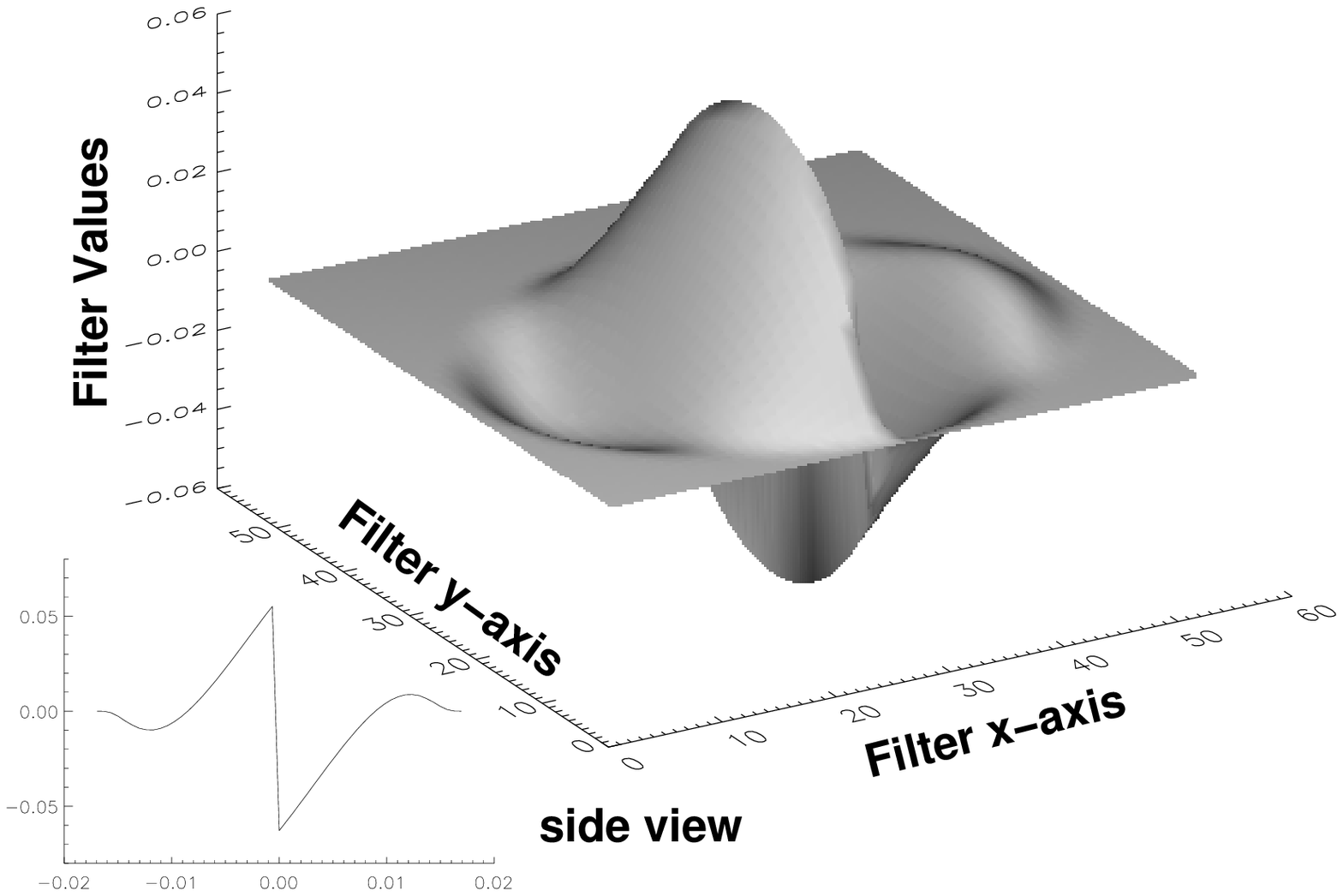}
\caption{A 3-D representation of the Edgefinder.  The z-axis represent
  filter values, and the x and y axis are the pixel numbers.  The
  right bottom inset is a plot of the cross section of the filter, with the x-axis
  being the distance from the filter center, and not the pixel number.
  \label{filter}}
\end{figure}

The filter values, $F(x,y)$, were given by,
\begin{equation}
 F(x,y) = N(x \mp A) \times \exp \left( \frac{-1}{1-r_n^2} \right)
\end{equation}
the minus sign applied to pixels with $x < 0$ and plus for $x > 0$,
$A$ was the normalized height of the filter, and $N$ was the total
normalization factor for the filter, discussed later.  The exponential
smoothing function ensured that the filter is smoothed and
compact. The normalized radius, $r_n$, is $r_n = \sqrt{x^2 + y^2}$.
This filter was designed to be insensitive to a constant background
value in the window, so that
\begin{equation}
\int F(x,y) dx dy = 0.
\end{equation}
The filter is also insensitive to gradients, 
\begin{equation}
\int x F(x,y) dx dy = 0,
\end{equation}
and the value of $A$ was adjusted to ensure this.  The Edgefinder is
depicted in Figure \ref{filter} as a shaded surface plot with the z
axis representing the filter value, $F(x,y)$. In addition, a plot of
the cross section of the filter, is also shown.

The filter could be rotated in the sky to detect strings of different
position angles with respect to the North-South Galactic alignment.
This was achieved by rotating through an angle $\alpha$ by altering
the $[x,y]$ values of a pixel to $[x',y']$ by the
following transformation,
\begin{equation}
\left ( \begin{array}{cc}
  x' \\
  y'
\end{array} \right) = \left ( \begin{array}{cc}
  x \\
  y
\end{array} \right) \left ( \begin{array}{cc}
  \cos \alpha & - \sin \alpha \\
  \sin \alpha &   \cos \alpha
\end{array} \right). 
\end{equation}
The filter values are then generated with $x'$ in place of $x$ and
$y'$ in place of $y$.  In most simulations, we ran the filter at 20
different $\alpha$ values equally spaced between 0 and 180
degrees. This represented a shift through one pixel at the edge of the
filter window, and should pick out edges aligned in all directions.

\begin{figure}[!h]
\plotone{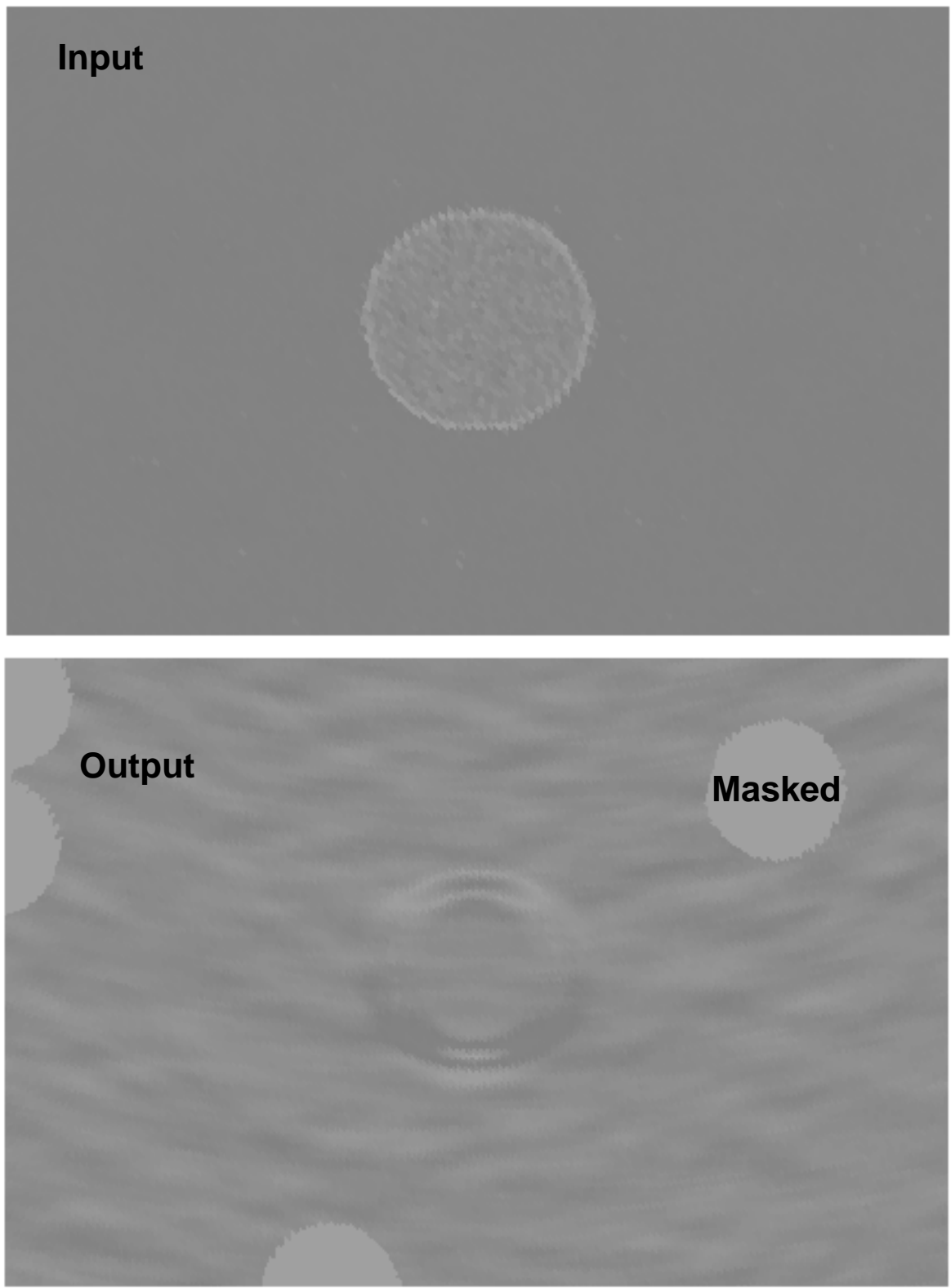}
\caption{Edgefinder values around a string horizon.  The Edgefinder is
  oriented North (up) to South (down).  The light blue circles are
  masked pixels in the WMAP data due to foreground sources. \label{edge_big}}
\end{figure}

The characteristic Edgefinder behavior around a string horizon is
depicted in Figure \ref{edge_big}.  In this figure, the top picture is
a simulated input temperature map, where the dark disk is the
simulated string horizon.  The disk edges are fuzzy due to the noise added
to the map (as a real string in the sky would be seen by WMAP).  The
bottom picture is the Edgefinder response to the input map.  The
Edgefinder was oriented North to South, and produced a ``hot'' signal
when encountering a rising edge, and a ``cool'' signal when
encountering a dropping edge.  If we rotated the filter window by
$180\degr$, we would get the opposite signal, meaning that the sign of
the EVs was not important, so we used absolute values in our
statistics. Running through 20 $\alpha$ angles from $0$ to $180 \degr$
really represented running through 40 $\alpha$ angles from $0$ to $360
\degr$.  

Visible on the same figure are pale circular disks which indicate
masked pixels in the WMAP data, usually where there were strong
foreground microwave sources.  These regions were not used in our
simulations.  We used the combined Kp0 and Kp2 masks (see
\citealt{ben03} for an explanation of the masks) which masked 1109593
out of 3145728 pixels (35.7\%).  If strings were partially blocked by
masked pixels, the Edgefinder can still detect them, but at a reduced
sensitivity.  We ran tests where strings were partially blocked by a
mask, with 5 different exposure levels from 100\% to 18\%, where 100\%
indicated an unblocked string.  These results are tabulated in Table
\ref{exposure}.  For strings which are more than 50\% exposed, there
were no differences in the peak EV compared to an unblocked string.  For
strings that are less than 50\% exposed, the peak EV dropped off
rapidly as the string centers moved behind the mask, but were
still detectable, their EVs dropping to around 15\% of an unblocked
string.  This test indicated that strings with centers located outside
of a masked area will be picked up normally by the Edgefinder, and
therefore the masks are not blocking more sky than their actual area.

The Edgefinder filter values, $F(x,y)$, were calibrated to give a
response of Edgefinder value, EV, of 1 for a 1 mK input signal.  For
example, if the input CMB map had all its southern hemisphere pixels
at 1 mK, and northern hemisphere pixels at 2 mK, the Edgefinder would
return EV = 1 for equatorial pixels, and EV = 0 at other pixels.  The
normalization constant for the $RAD = 1\degr$ WMAP Edgefinder was $N =
0.33$. Due to fluctuations in the normalization and the excess value
collection of the filter, the actual response varied between 0.97 and
1.08, but averaged to 1.0 across the equator. Strictly speaking, the
EV is unit-less, but for the sake of clarity, we will sometimes give
it units of mK.  Unless specifically mentioned, all temperature units
in this paper are in milliKelvin.

We created a set of calibrators to verify the gain of the Edgefinder.
The calibrating set contained edges, or string horizons, of various
temperatures ranging from $\Delta T/T = 1.1$ mK to $\Delta T/T = 1$
nK.  The results of the calibration are in Table \ref{cali_val}. In
most cases, the EV was the same magnitude as the input edge to within
2\%; we are therefore confident that the Edgefinder had a linear
response to the input string horizon over the range of pertinent input
temperatures.

\begin{figure}
\plotone{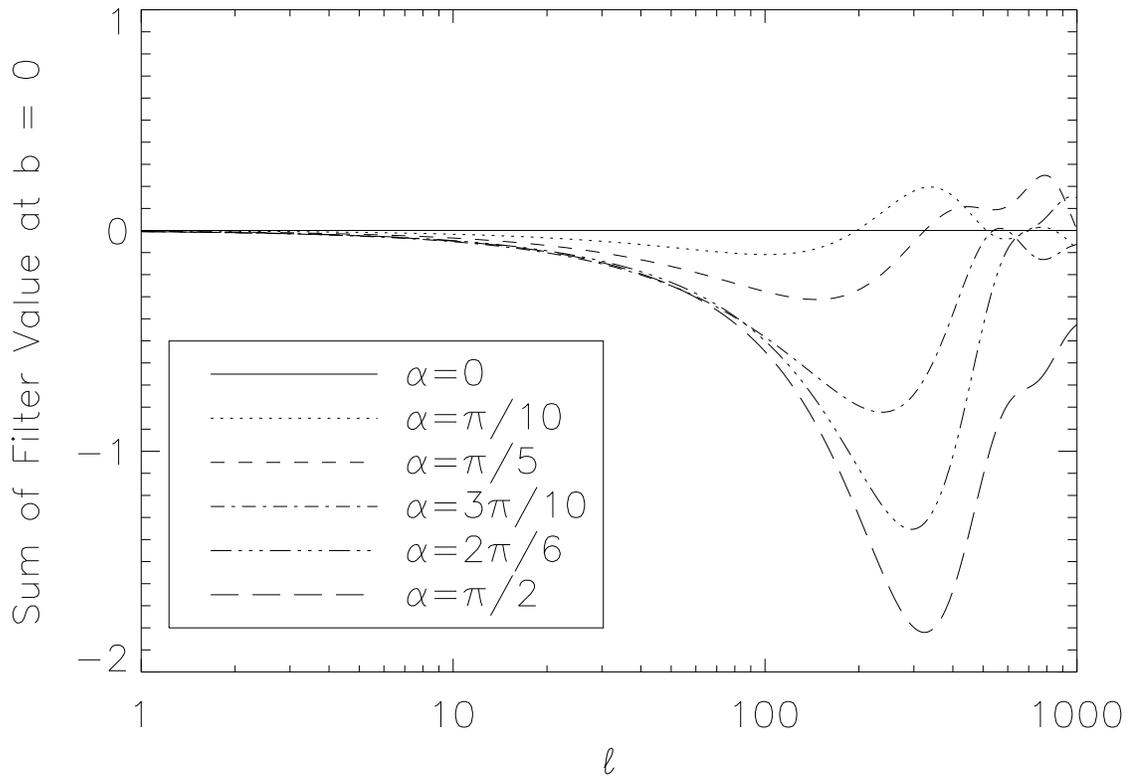}
\caption{The response function of the Edgefinder Filter.  Note the
  peak at $\ell \sim 200$, around where the first Doppler peak of the CMB
  anisotropy angular power spectrum occurs. \label{filt_res}}
\end{figure}

The size of the Edgefinder was chosen to match the most likely string
size that we may be able to find.  For WMAP, this was a problem since 
this scale was also where the CMB Gaussian anisotropy signals were the
strongest.  The response of the filter to $\ell$ values from 1 to 1000
is plotted in Figure \ref{filt_res} for representative $\alpha$
angles.  The largest filter response was around $\ell = 200$; the WMAP
data indicated that the first Doppler peak of the CMB anisotropy
angular power spectrum is located at $\ell \sim 220$.  This was
undesirable as we wanted to insensitive to as much much of the
Gaussian signal as possible so we could focus on the non-Gaussian
signals.

One easy solution would be to make the filter smaller. However, this
is not feasible with WMAP data. At $RAD = 1\degr$, there were 273 pixels in
the filter; at $RAD = 0.5\degr$, there would be only 69 pixels in the
filter, spanning less than 10 pixels. This resolution was too
coarse to run the Edgefinder.  For now, we report results of our good
enough filter while we work on ways to eliminate the response at $\ell
\sim 200$.  One certain method is to wait until PLANCK data is
available, and with the improved resolution, we will get ~270 pixels
in the filter at $RAD = 0.5$.  This is discussed in Section 6.

\subsection{Simulated Maps}

We knew from the calibrator set described in the previous section that
the Edgefinder had a linear response to the input edge.  We next needed to
find out how the Edgefinder responded to noisy maps like the ones from
WMAP. We anticipated that the Edgefinder would not be able to detect
strings when the string horizon signals became swamped with noise.  In
order to find this limit, we produced simulated maps to quantify the
behavior of the Edgefinder.  Once we understood how the Edgefinder
responded to maps containing strings of known magnitudes, we could then
set detection limits of the Edgefinder for the WMAP data. Because they
are so vital, we describe in detail how we generated the simulated maps:

1. Coefficients, $C_{\ell}$, of the CMB angular power spectrum were
generated by CMBFAST using the cosmological constants derived from the
WMAP experiment, a $\Lambda$CDM cosmology, given by:
$\ell_{max} =      1500$, 
$\kappa_{max} =     3000$,  
$\Omega_b=          0.044$,  
$\Omega_c=          0.218$, 
$\Omega_{\lambda}=  0.738$,  
$H_o=               71.6$,   
$T_{CMB}=           2.725$, 
$Y_{He}=            0.24$,  
$N_{\nu (massless)}=3.04$,  
$\tau_{LSS}=        0.099$, and 
$n  =                0.955$ \citep{ben03}.

2. We used HEALPix associated software SYNFAST to generate a CMB map
which matched the $C_{\ell}$ generated in Step 1.  SYNFAST generates
random Gaussian fields on a sphere based on the input power spectrum
(the $C_{\ell}$'s).

3. We added string horizons at various temperatures, $T_s$, into the
map by adding the value $T_s$ to circular regions within a string radius 
$R_s$.

4. We used HEALPix associated software SMOOTHING to convolve three
beams with the map made in step 3. A W-Band map was made by convolving
a 13 arc-minute Gaussian beam with the $a_{lm}$'s of the temperature
map created at Step 3, and the map regenerated with the new
$a_{lm}$'s.  A V-Band map was made from the convolution of a 21
arc-minute beam, and a Q-Band map from a 32 arc-minute beam.  For each
map generated in Step 3, three maps convolved with beams appropriate
for the three WMAP bands were made.

5. The 3, W-, V-, and Q-Band, maps were averaged to get the final
SMOOTHed map.  Noise was then added to the map with the following
prescription: the $\sigma_n$ for the noise of each pixel was generated
by combining the noise characteristics of the 3 WMAP bands,
\begin{equation}
\sigma_n = \frac{\sqrt{ \sigma_Q^2 + \sigma_V^2 + \sigma_W^2} }{3}
\end{equation}
where $\sigma_Q$ is the pixel noise in the $Q$ band,
\begin{eqnarray}
\sigma_{Q} & = & \frac{\sigma_{Q,0}}{\sqrt{N_{Q,obs}}} \\ 
\sigma_{V} & = & \frac{\sigma_{V,0}}{\sqrt{N_{V,obs}}} \\ 
\sigma_{W} & = & \frac{\sigma_{W,0}}{\sqrt{N_{W,obs}}} 
\end{eqnarray}
where $N_{Q,obs}$ is the number of times the pixel had been observed
by the WMAP satellite in the $Q$ band; $\sigma_{Q,0} = 2.211,
\sigma_{V,0} = 3.112, \sigma_{W,0} = 6.498$ are the noise weights
given by the WMAP team.  The noise values were generated by a Gaussian
random number generator with $\sigma = \sigma_n$.  The reason for this
noise prescription, and for the convolution with three beams in Step
4, was that the data we fed to the Edgefinder are a composite of data
from the three WMAP bands Q, V and W.  Due to the SMOOTHING and the
added noise, the final string horizon temperature was at a slightly
different temperature than the initial $T_s$. We sometimes also
report, for comparison, the average temperature of the final input
horizon, which we designated $T_f$.

\subsection{Edgefinder Limits}

In all, 82 input simulated maps were made according to the
prescription given in Section 4.1.  Out of these, 15 of the input maps
were baseline maps, where no string horizons were inserted.  We called
the EV output of these maps the No-String sets.  The maximum EV of a
the No-String sets indicated the maximum EV due to background signals
(i.e., not from the strings).  In other words, the No-String sets
produced the noise limit; we considered anything above this limit to
be signal.  Both the CMBFAST generated CMB signal and noise are
basically Gaussian random variables, so we needed to have a range of
No-String sets in order to ensure we have proper coverage of the
possible noise values.  To this end, we created 15 No-String input
maps.

We then created the 67 ``Stringy'' maps, which contained different
number of string horizons with various sizes and temperatures.  The
input string horizon temperatures ranged from $T_s$ = 1.0 mK to 0.0001
mK, the inserted string radius, $R_s$, ranged from 1.0 to 4.0 degrees.
Most maps were made with one inserted string, but one map contained as
many as 60.  The Stringy maps are further divided into single and
Multi maps, where Multi maps had more than 1 inserted string horizons.
There are 16 Multi maps.

\begin{figure}
\plotone{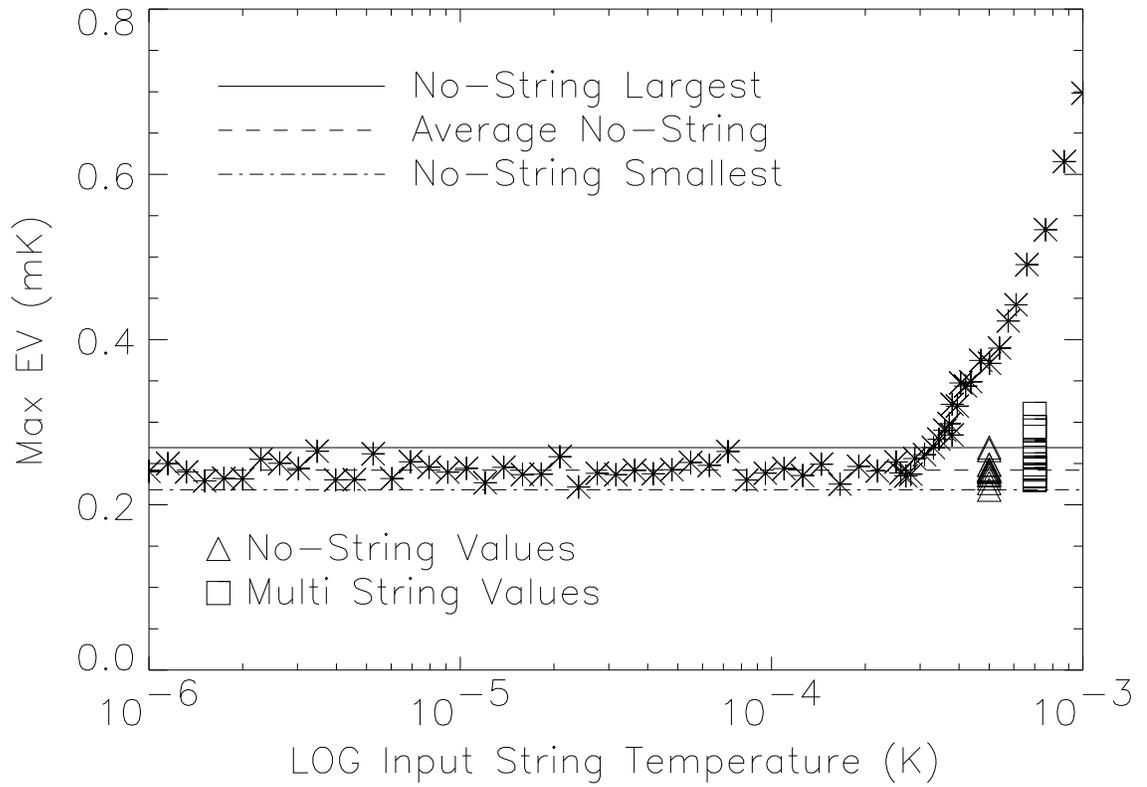}
\caption{Plot of the input string temperature $T_f$ vs. the maximum EV
  of the set. \label{max_ev}}
\end{figure}

\begin{figure}[h]
\plotone{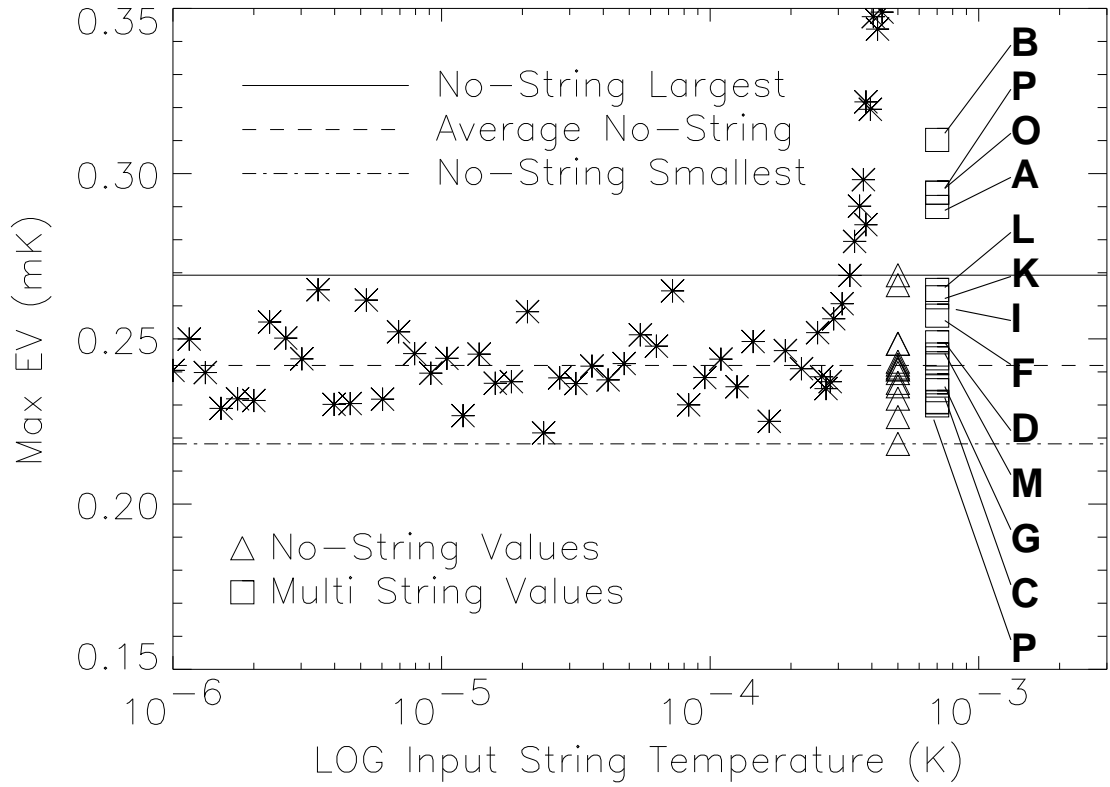}
\caption{Detail of the plot of the input string temperature $T_f$
  vs. the maximum EV. Also plotted are the multi-string
  set data as well as limits from the No-String sets.\label{max_ev_big}}
\end{figure}

Comparison between the maximum EV (Max EV) of the Stringy and
No-String sets will yield the input string value that gives a Max EV
above the No-String limit.  The No-String limit is given by the
largest value of the Max EVs of the No-String sets.  The average of
the maxima of the 15 No-String sets was EV = 0.242 mK; the largest of
the 15 sets was EV = 0.269 mK.  We have plotted the Max EV of all
single Stringy sets against the input string temperature, $T_s$ in
Figure \ref{max_ev}.  The No-String limits are plotted as the dashed
and dotted horizontal lines.  A blow up of the region around the
No-String limit is in Figure \ref{max_ev_big}.  The data indicates
that an input string horizon of 0.345 mK and cooler yielded similar
Max EVs as the No-String sets.  We can consider T = 0.354 mK as one of
the limits to the sensitivity of the Edgefinder.  If we took this
limit to be wakes formed by strings traveling at $c/\sqrt{2}$ (the
mean absolute velocity of strings from numerical simulations),
according to Equation \ref{string_t}, we would have limits for the
cosmic string at $G\mu \la 1.37 \times 10^{-5}$.

This is the crudest but most robust method of obtaining the
sensitivity limit of the Edgefinder.  The problem posed by the
Edgefinder was that we were looking for very small non-Gaussian
features in a largely Gaussian data set.  For an input string of $R_s
= 2\degr$, the number of pixels whose filter window contained the edge
constituted 0.0087\% of the total pixels.  The binning of the EV set
data resulted in the non-Gaussian signatures occurring in the edges of
the histogram.  To improve upon the Max EV limits, we needed a set of
descriptive statistics that could pick out small non-Gaussian signals
at the outer edges of a Gaussian function.  We found the Edgeworth
Series to be suitable for our purposes.

\subsection{Edgefinder Limits from Edgeworth Coefficients}

We have binned the EVs such that the cosmic string signals are in
the wings of the distribution.  Since the CMB background is Gaussian,
and we are looking for small deviations from a Gaussian, it is
natural to consider using a series involving the Gaussian function,
$\alpha(x) = \exp(-x^2/2)/\sqrt{2\pi}$, and its derivatives to fit our 
distribution, $dF$.  Such a series would have the form,
\begin{equation}
\label{gc-roughs}
f(x) \sim \sum_{j=0}^{\infty} c_j H_j \alpha(x)
\end{equation}
where $c_js$ are coefficients to the $H_js$, which are the Hermite
polynomials. The $j^{th}$ Hermite polynomial is the polynomial
resulting from the $j^{th}$ derivation of the Gaussian function
$\alpha(x)$.  The Hermite polynomials have a generating function,
\begin{equation}
H_j(x) = (-1)^j e^{x^2} \frac{d^j}{dx^j}e^{-x^2}.
\end{equation}

The series of the form given in Equation \ref{gc-roughs} that we have
chosen to use is the Edgeworth series.  The Edgeworth series is an
asymptotic expansion of a distribution as a function of the
$\alpha(x)$ and its derivatives.  A full derivation of the Edgeworth
series can be found in \citet{ken87}, and discussions relevant to
astrophysics can be found in e.g. \citet{jus95} and \citet{bli98}.

The general form of the Edgeworth series is given by,
\begin{equation}
f(x) = \left (E_1+E_2 H_3+E_3 H_4+E_4 H_5+E_5 H_6+...\right )
\alpha(x) 
\end{equation}
The coefficients of the series, $E_n$, are given in terms of the
cumulants, $\kappa$, of the distribution, $dF$.  The $j^{th}$ cumulant
is a linear combination of the $j^{th}$ and lower order moments, $\mu_j$. We
obtain the cumulants of our distribution from calculating the moments
of the distribution, $dF$, by,
\begin{equation}
\mu_j = \int_{-\infty}^{\infty} (x - \bar{x})^j dF
\end{equation}
where $\bar{x}$ is the mean of the distribution.  The relationship
between cumulants and moments is given by,
\begin{equation}
\mu_r = \sum_{j=1}^{r} \left( \begin{array}{c} r-1\\
  j-1\end{array} \right ) \mu_{r-j}\kappa_j  
\end{equation}
where the bracket is the binomial bracket. The first seven non-zero Edgeworth
series coefficient are,
\begin{eqnarray}
E_1 & = & 1 \nonumber \\
E_2 & = & \kappa_3/6 \nonumber \\
E_3 & = & \kappa_4/24 \nonumber \\
E_4 & = & \kappa_5/120 \nonumber \\
E_5 & = & (\kappa_6 + 10\kappa_3^2)/720 \nonumber \\
E_6 & = & (\kappa_7 + 35\kappa_4\kappa_3)/5040 \nonumber \\
E_7 & = & (\kappa_8+56\kappa_5\kappa_3+35\kappa_4^2)/40320. \nonumber 
\end{eqnarray}

We tested the Edgeworth Series on the histogram of a perfect Gaussian
distribution, for which we knew all $E_n$ coefficients should be zero.
We found that the biggest effect on the coefficients was how
quickly the edges of the Gaussian function went to zero.  For example,
in two trial runs we used 1 million random numbers to represent a
normal Gaussian distribution, with zero mean and unit variance.  The
one run where we allowed the x variable to go to $x = 16$, generated
zeros at a level of $10^{-17}$.  The other run where we only allowed
x to go to $x = 3$, had zeros on the level of $10^{-8}$.

This meant that we had to choose the same set of bins for every EV set
instead of binning the data with bin-size set by the maximum and
minimum of each distribution.  We experimented with the number of bins
to see what our bin-size should be.  With too few bins, the
non-Gaussian aspects of the distribution were hidden.  With too many
bins, the residuals ($\sum (data - fit)$) increased from errors at the
wings of the distribution, and since this was exactly the location of
the non-Gaussian signals we wished to detect, we had to be cautious.
With the number of bins from 500 and 1000, the variation in the
coefficients generated were within 10\%.  The optimal number of bins
turned out to be 574, which struck a balance between generating a good
zero value for the perfect Gaussian case, as well as having enough
room at the wings for the non-Gaussian signals.  We binned the
absolute value of the EV set for each of the simulated maps into 574
bins, with the uppermost bin at 0.3 mK.  The upper bin was chosen
because maps containing EV greater than 0.3 mK were well above the Max
EV of the No-String set and therefore were already known to contain
strings.  Since we used the absolute value of the EVs, we mirrored the
histogram across the y-axis to create a negative half for the
distribution; the whole distribution was then normalized.

\begin{figure}
\plotone{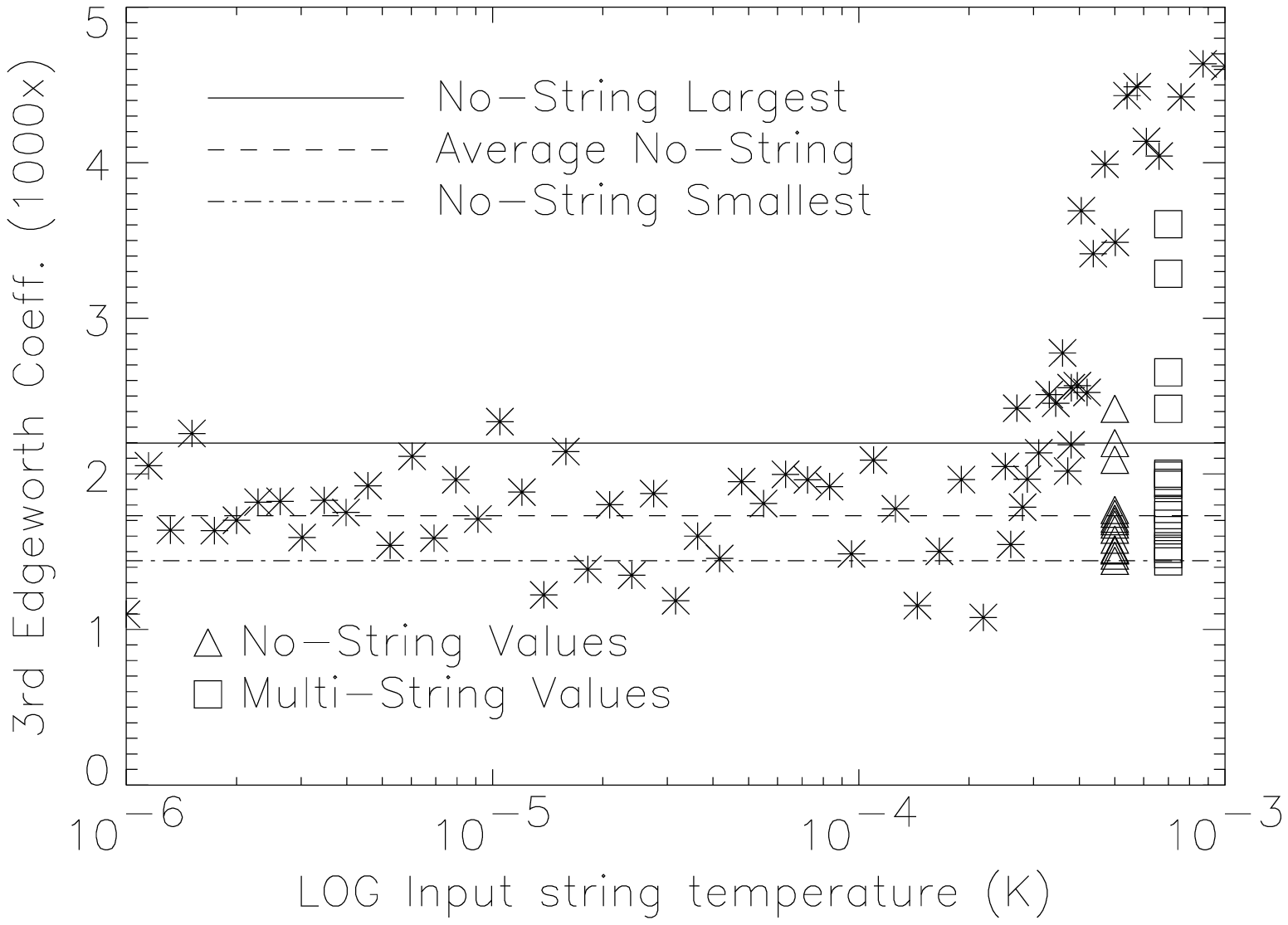}
\caption{Plot of the input string temperature $T_f$ vs. the 3rd
  Edgeworth coefficient. \label{E_3}}
\end{figure}

\begin{figure}
\plotone{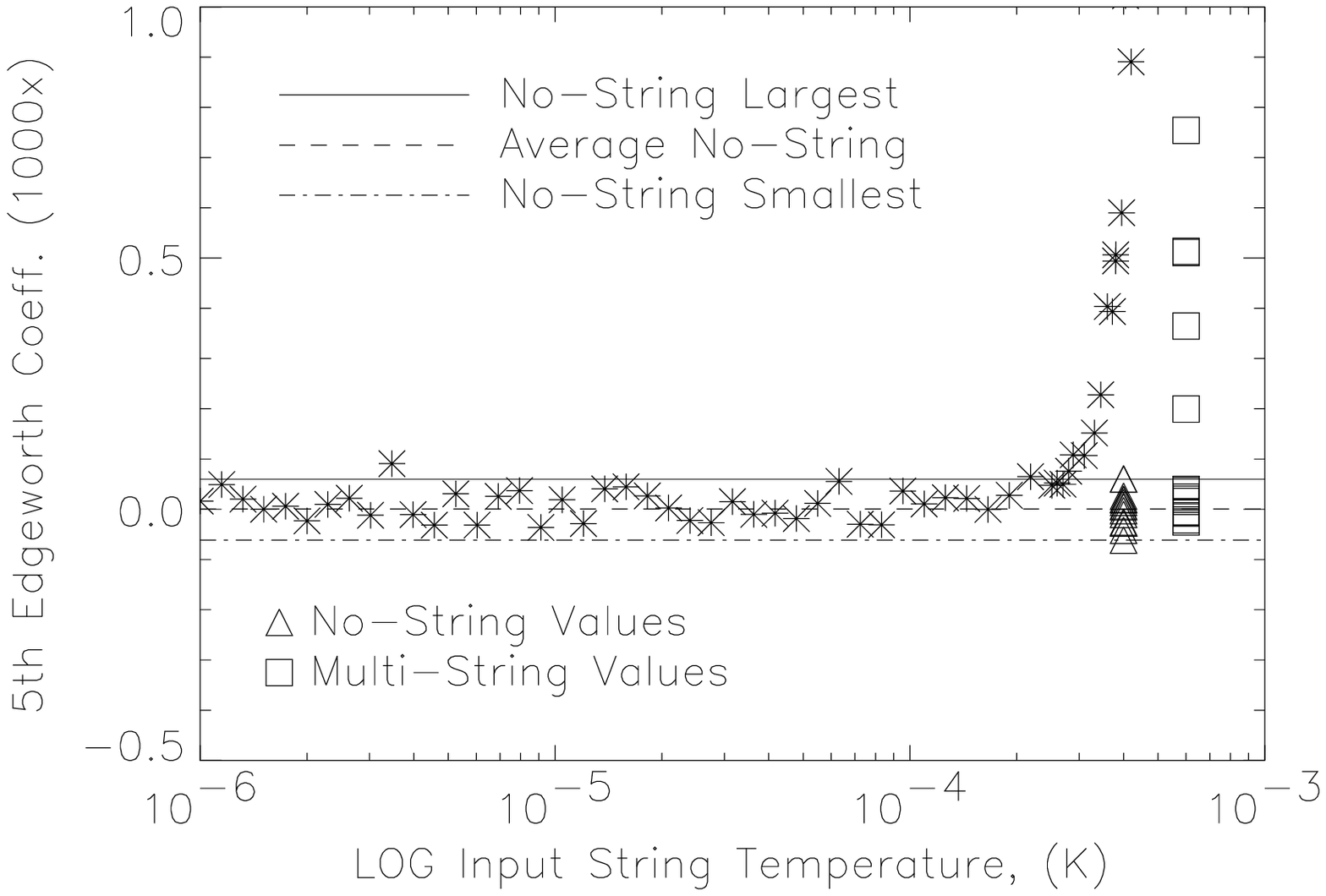}
\caption{Plot of the input string temperature $T_f$ vs. the 5th
  Edgeworth coefficient. \label{E_5}}
\end{figure}

Due to the fact that we reflected the distribution across the y-axis,
the histograms were completely symmetrical. This meant that the odd
moments (even cumulants) were zero.  Thus, we only needed to look at
the odd Edgeworth coefficients; $E_n$ where $n$ is 3, 5, or 7.  For
our purposes, the 7$^{th}$ Edgeworth coefficient, $E_7$, was a good
discriminator of whether a map contained a cosmic string horizon.  We
have included plots of the other coefficients, $E_3$ and $E_5$, in
Figure \ref{E_3} and Figure \ref{E_5} respectively; note that the
y-axis of Figure \ref{E_3} is on a log scale to include all data
points.  Compared to the $E_7$ coefficient, $E_3$ did not pick out as
many points, so at best, it can used as corroborating evidence.  The
$E_5$ coefficient was similar to $E_3$, but it included more points.

Out of the 15 sets of No-String maps, the $E_7$ coefficients ranged
from $6.51 \times 10^{-6}$ to $-3.92 \times 10^{-6}$.  Inspection of
$E_7$ and other coefficients showed that there was a clear trend: the
Stringy sets which contained hot string horizons produced bigger
non-Gaussian tails that were picked up by $E_7$.  The fact that the
$E_7$ coefficients of the No-String set were distributed around zero
tells us that a small $E_7$ indicated a lack of non-Gaussian features
like those produced by a string horizon.  It also indicated the
magnitude at which the $E_7$ coefficient can be considered noise.
This was corroborated by the fact that the higher the $T_f$, the
larger and more positive the $E_7$ coefficient.  Therefore, we could
use the No-String $E_7$ values as a discriminator between maps with
strings and maps without strings. We took the second largest No-String
(14th out of 15) $E_7$ value as our limit of the Edgefinder's single
string sensitivity, at EV = $5.06 \times10^{-3}$ mK.

\begin{figure}
\plotone{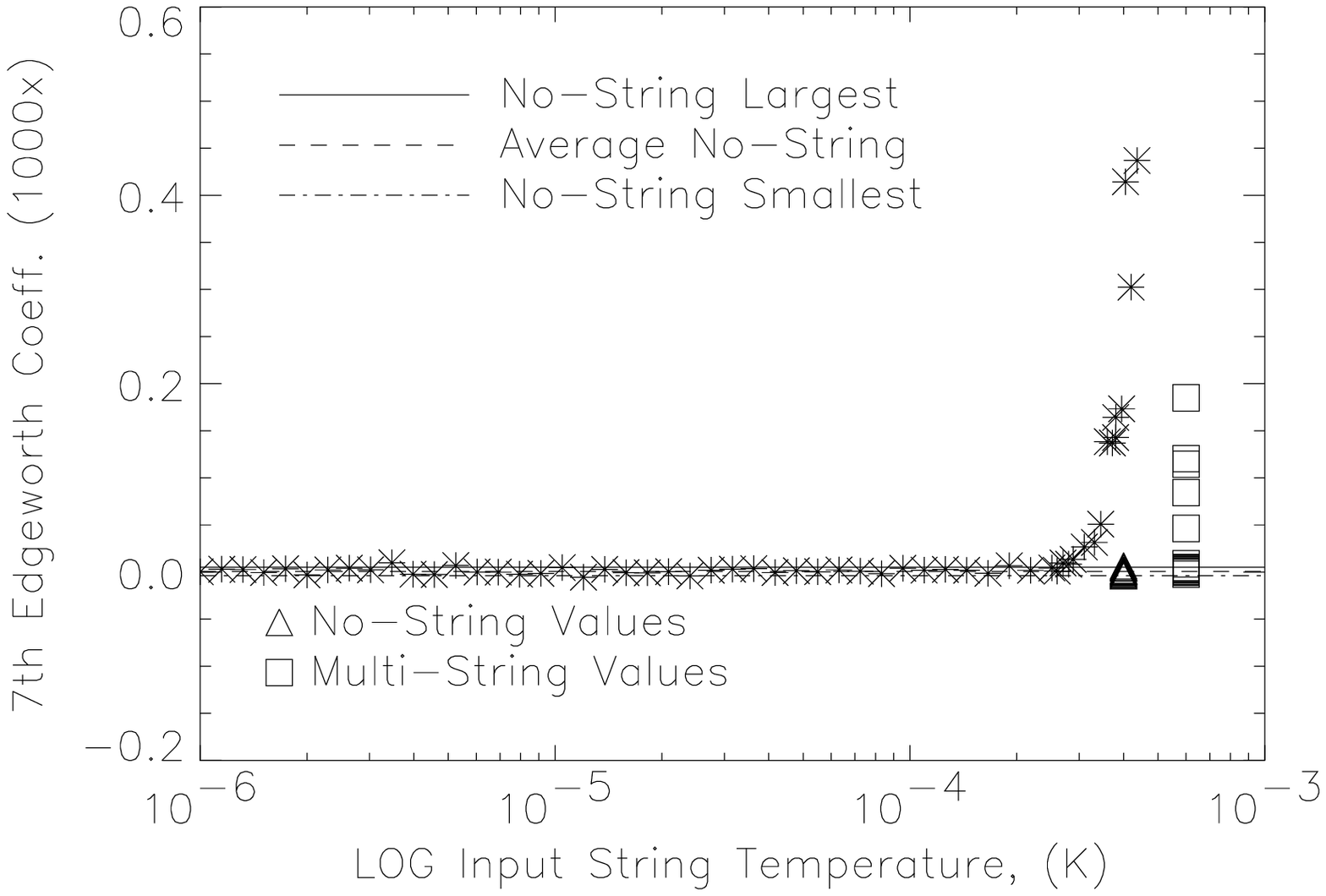}
\caption{Plot of the input string temperature $T_f$ vs. the 7th
  Edgeworth coefficient. \label{E_7}}
\end{figure}

\begin{figure}
\plotone{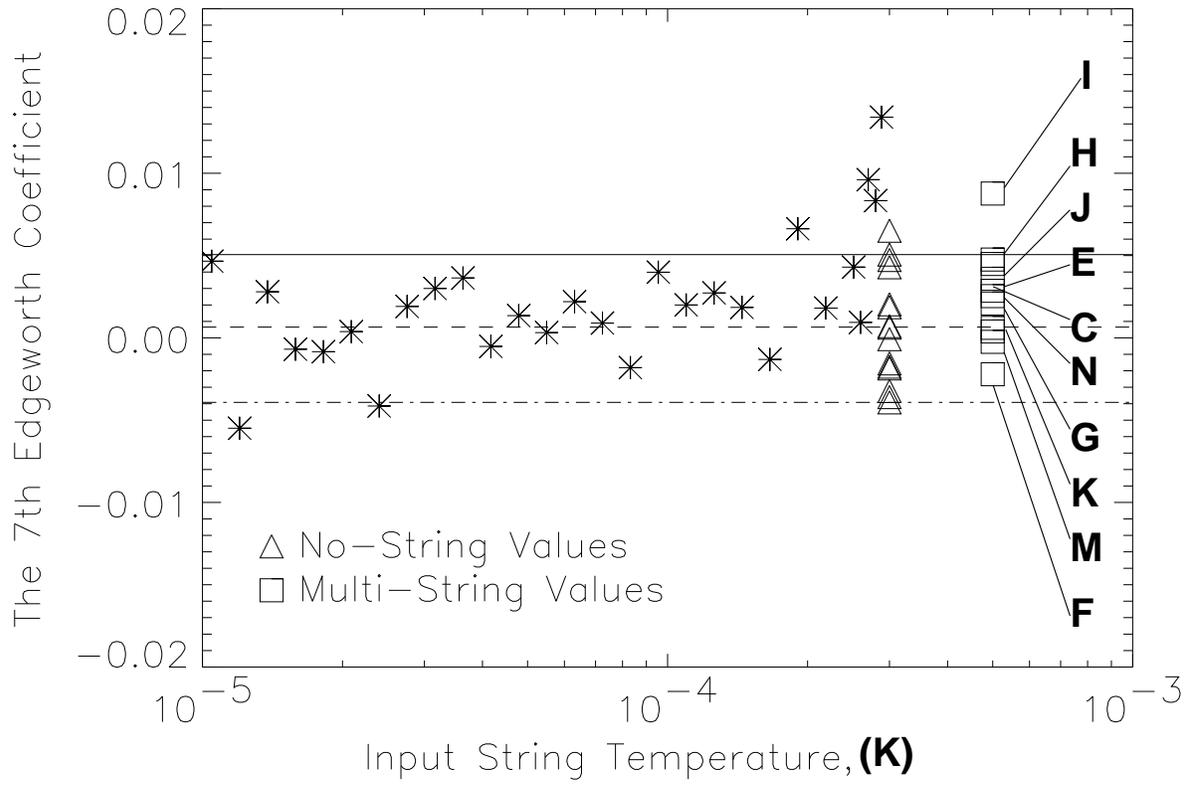}
\caption{Detail of the plot of the input string temperature $T_f$
  vs. the 7th Edgeworth coefficient. Also plotted are the multi-string
  set data as well as limits from the No-String sets.\label{E_7_big}}
\end{figure}

The type of map that produced the smallest $E_7$ signals were maps
with only one string horizon inserted.  We made 66 maps of single
string with input temperatures ranging from 1 mK to 1 $\mu$K. The
$E_7$ coefficients and input string horizons for all sets are graphed
in Figure \ref{E_7}.  It shows clearly that a large and positive value
of $E_7$ is indicative of the presence of hot string horizons.  Figure
\ref{E_7_big} is a blow up of the region around the No-String limit.
String horizons with $T_s > 0.27$ mK were clearly above threshold
$E_7$ value.  At $T_s = 0.260$ mK we encountered the first string
horizon with $E_7$ below the threshold.  We therefore set $T_f = 0.27$
mK as our 100\% confidence level of string detection.  Examining
Figure \ref{E_7_big}, we can see that the $E_7$ values mostly fall
within the boundary defined by the maximum No-String and minimum
No-String $E_7$, with a few peeking above the threshold.  The scatter
in the data is very constant.  Therefore the limit of our detection is
firm at 0.27 mK; all input strings cooler this threshold looks like
noise to the Edgefinder.  For a cosmic string moving at mean
simulation velocity, $\beta = 1/\sqrt{2}$ this corresponds to a $G\mu
= 1.07 \times 10^{-5}$ string.

\subsection{Multiple String Maps}

17 simulated maps were made with multiple string horizons inserted
in the map, ranging from 1 to 60 horizons and $R_s$ ranging from 1 to
4$\degr$.  The number of string horizons and their sizes are
summarized in Table \ref{multi_s}.  The purpose of the multiple string
sets were twofold, first to determine if multiple strings had a
similar effect on the Edgeworth coefficients as the single strings,
and if so, determine the number of weak strings needed to generate
an $E_7$ signal above the No-String discriminant.  The Max EV and the
$E_7$ coefficients for the multiple sets are plotted as letters A to
P, according to their set name in Table \ref{multi_s}, in Figures
\ref{max_ev_big} and Figure \ref{E_7_big}, respectively.  The bold
items in the table represent sets with $E_7$ coefficients larger than
the No-String limit.

From the results of Sets A to C, we can see that the Max EV of
multiple string sets are generated by individual strings, and
therefore from the Max EV alone, we cannot tell if there are more than
one hot string in the data.  However, the $E_7$ coefficient of
multiple strings are cumulative.  This means that in a situation
where we have a small Max EV, and a large $E_7$, we know that the sky
contains multiple strings. 

However, looking at the results of Sets D, E, and F, we can see that
if the strings are cooler than the detection threshold, their $E_7$
signatures are small enough that a few strings (less than 10) will not
show detection.  It takes quite a few cool strings for the cumulative
effect to show up.  For very cool strings ($T_s \sim 0.15$mK), it takes 20
input strings for the $E_7$ coefficient to be above the threshold.

The size of the string has some effect on the coefficients, but it is
secondary to the strength of the edge.  Compare Sets D to J and G to
K; these sets have similar number of input strings, and similar $T_s$,
but Sets J and K have twice the $R_s$, which means 4 times the number
of pixels.  The $E_7$ coefficients of the sets are similar and show little
trend reflecting the change in inserted string size.  Furthermore,
consider Set L, where the inserted string had $R_s = 1$ degree; this
set had an $E_7$ coefficient significantly above the detection
threshold.  Both these phenomenon point to the fact that the
Edgefinder is most sensitive to the temperature jump at the edge of a
string horizon, rather than the size of the string.

The results of the multiple string sets shows that the Edgefinder
limit is firmly set at $T_s \sim 0.27$ mK.  We can detect cooler
strings, but they need to be numerous: looking at sets M and N shows
that we need more than $\sim$10 strings of 0.15 mK for a detection.
For very cool strings at  $T_s < 0.03$ mK, the data is very noisy.
The multiple strings sets indicated that the most important criterion
for string detection is the temperature of the input string.  If the
string is above the detection threshold, within a reasonable range of
sizes, it will be picked up by the Edgefinder. 

\section{WMAP Results}

The WMAP data we ran through the Edgefinder is an average of the data
in three of the WMAP bands: Q, V and W.  We called this the QVW
composite map.  Pertinent statistics of the WMAP QVW composite map are
in Table \ref{WMAP_stats}, including the Max EV and its $E_7$
coefficient.  The resulting EV set was binned in the same manner as
the simulated maps.  Comparing the QVW composite map and the No-String
sets, the statistics are similar.

Looking at the results of the WMAP composite QVW data, the maximum EV
is below the threshold set from the max EV of the No-String set, and
is in fact, below the maximum EV of half of the No-String simulated
data sets.  This means that there are no hot string horizons in the
WMAP data at the level of noise set by the No-String sets.  The $E_7$
coefficient leaves no doubt that there are no non-Gaussian signatures
in the tail of the distribution of the WMAP data produced by string
horizons similar to those we inserted into the simulated maps.  We can
confidently say that the Edgefinder did not find any evidence of
string wakes in the CMB data measured by WMAP, to the single string
limit of $G\mu < 1.07  \times 10^{-5}$.  The 20 string limit of
less 0.15 mK strings gives $G\mu < 5.97 \times 10^{-6}$.

We mention again the fact that in the actual WMAP data, due to
Galactic and foreground contamination, about 1/3 of the pixels were
masked.  We have reproduced this masking in our simulated data so that
we have the same number of pixels per map as the
actual WMAP data.  We caution that any strings hidden behind these
masked pixels would not be picked up by the Edgefinder.

\subsection{WMAP 2nd Year Simulation}

We have performed the analysis necessary in anticipation of the WMAP
second year data release.  All of the analysis were done in the same
manner as the first year data, with the exception that the noise is now
$1/\sqrt{2}$ time the noise of the first year data.  The second year
data has limits at max EV = 0.283 mK and $E_7$ = 0.257 mK.  The $E_7$
threshold yields a single string limit of $G\mu < 1.02 \times
10^{-5}$.  Most of the noise in our string search is due to the first
acoustic peak in the CMB angular power spectrum; the Edgefinder is
cosmic variance limited, so reduced radiometer noise has little effect. 

\section{A Cosmic String Candidate?}

Sazhin et al.(2003) reported a discovery of an object which contains
two sources of identical isophotes, color, and fitted 2-D light
profiles in the Osservatorio Astronomico di Campodimonte Deep Field
(OACDF).  In addition, spectra of the sources are identical with a
confidence level higher than 99\%.  Morphological arguments led them
to propose that this object is a background galaxy lensed by a cosmic
string.  They have named this object the Campodimonte-Sternberg-Lens
candidate 1, or CSL-1.  The red-shift of both sources in the object is
$0.46 \pm 0.008$; the separation of the two sources is 2 \arcsec.

Precise finder charts for the object were not available. We found this
object by visually inspecting the OACDF deep field and comparing it to
the Palomar All Sky Survey plates.  We found CSL-1 to be located at
(J2000) RA 12:23:30.72, Dec -12:38:57.8.  There may be some small
uncertainty about the location of CSL-1. We examine the WMAP data at
this location to see if the Edgefinder can detect a string.  As WMAP
data had a resolution of 0.23 degrees, by including the four nearest
pixels to that coordinate, we believe we have covered this object in
our search.

\subsection{String Search}
From the image separation, we can derive a string mass per unit
length, $\mu$.  In a flat universe where $\Omega_{matter} = 1$, from Gott
(1985), we find that the image separation $\Delta \theta$ of lensing
by a cosmic string is related to $D$, the deficit angle of the conical
space around the string, by,
\begin{equation}
\Delta \theta = D \left[ \cos \alpha - \frac{1-(1 + z_s)^{-1/2}}{1-(1
    + z_g)^{-1/2}} \right]
\end{equation}
where $z_s$ is the red-shift of the string, $z_g = 0.48$ is the red-shift of
the background object (galaxy), and $\alpha$ is the angle of the
straight string with respect to the plane of the sky.  If we assume
that the string is in the plane of the sky, which means $\alpha = 0$,
there are two limiting results: for $z_s = 0$, we find $\Delta \theta
= D$; for $z_s = 0.4$, $\Delta \theta = 0.102D$.

For a flat universe, $D$, is related to the string mass per unit
length, $G\mu$ by,
\begin{equation}
D = 8\pi G \mu.
\end{equation}
Therefore, with a maximum $D$ of $9.7 \times 10^{-6}$ radians, we get
$G \mu = 3.86 \times 10^{-7}$.  A string will cause a jump in the
value of the temperature of the CMB due to
Doppler shifting.  The change in temperature is given by Equation
\ref{string_t}.

We have processed WMAP data at the position of the CSL-1 object and
compared the Edgefinder values to Edgefinder values of regions with
similar sky coverage and galactic latitude.  The CSL-1 object was
small enough that it was within one WMAP pixel (on pixel 968549).
However, we include results from the surrounding 4 pixels in the event
that we have misjudged the position of CSL-1.  In addition, the
alignment of the two images contains some uncertainty, thus we also
ran the Edgefinder along two separate position angles, $\alpha = 0$
and $\alpha = \pi/20$.  The results are in Table \ref{tbl-1}.

The Edgefinder values of the 4 pixels are all above the 95th
percentile, especially pixels 968548 and 968527, which have very high
E.V.'s above 99th percentile. However, these high percentiles can be
misleading. First, they are all significantly under the No-String
detection limit.  Second, as an upper limit, if we allow the WMAP
temperature at those pixels to be entirely caused by the presence of a
string, we can say that the EV is exactly the $\Delta T$ due to the
string motion, due to the Edgefinder gain being 1.  With the $G\mu$
given above, this means for an E.V. = 0.08686 mK, the string needs
$\beta\gamma \ga 3.3$, or $v = 0.957 c$, to account for the
temperature jump.  The ranges of string velocity for the E.V. in Table
\ref{tbl-1} is from $v = 0.941 c$ to $v = 0.979 c$.  These high string
velocities makes the case for the existence of cosmic strings at this
location more unlikely, as the rms string velocity is $v \sim 0.7c$.
We therefore cannot say that we have a significant detection of a cosmic
string at the location of CSL-1 in the WMAP data.

\section{Simulated PLANCK Results}

Looking forward, we also made simulated maps of CMB data that will be
gathered by European Space Agency's PLANCK satellite. PLANCK is
scheduled to launch in 2007.  Most of PLANCK is still
being constructed, so we do not have data on the noise
characteristics of PLANCK.  What we do know is that PLANCK will have a
5 arc-minute resolution, with a temperature sensitivity of $4 \times
10^{-6}$ K (more information is available from the PLANCK
website\footnote{\url{http:/www.rssd.esa.int/index.php?project=PLANCK}}).
We expect PLANCK will have about 1/10 the noise of WMAP.
For our simulated PLANCK maps, we followed the same procedure outlined
in Section 4, except in step 5, where for the added noise, we used 1/10 the
average WMAP noise value for the $\sigma_n$ of each PLANCK pixel.  In
the HEALPix scheme, the improved PLANCK resolution means having 12
million pixels in the sky, a 4 fold increase on WMAP. Running each
simulated map at 20 $\alpha$ angles became computationally infeasible.
Therefore, we only ran the PLANCK maps through 6 different $\alpha$
angles, spaced equally between $0$ and $180 \degr$.

\begin{figure}
\plotone{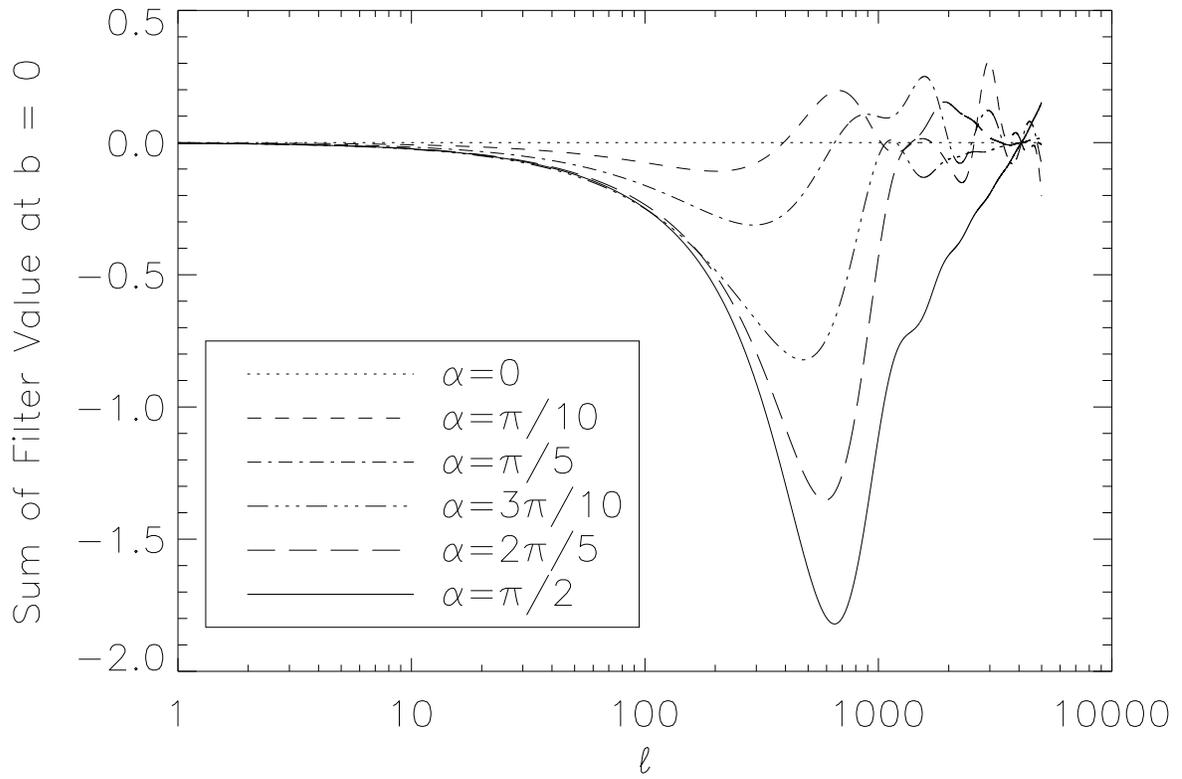}
\caption{Transfer function for the PLANCK simulation; the filter had
  $RAD = 0.5$. \label{p_trans_fn}}
\end{figure}

The increased resolution of the PLANCK satellite will allow us to
bypass the problems of being sensitive to the first Doppler peak at
$\ell \sim 200$.  With the PLANCK Edgefinder at $RAD = 0.5\degr$, we still
had $\sim$ 270 pixels in the filter window, which was comparable to
the number of pixels in the WMAP Edgefinder window of $RAD = 1.0\degr$.
The transfer function for the $RAD = 0.5\degr$ PLANCK Edgefinder filter is
plotted in Figure \ref{p_trans_fn}.  For the $RAD = 0.5\degr$
Edgefinder, the peak filter sensitivity is at $\ell \sim 500$, well
away from the first Doppler peak.  We performed the bulk of our
simulation using the $RAD = 0.5\degr$ PLANCK Edgefinder.  At this
resolution, the PLANCK filter had a similar number of pixels as the
WMAP $RAD= 1\degr$ filter, so the gain from the filter was comparable.

As with the WMAP set, we created calibrators for the PLANCK Edgefinder
and tuned the filter to return EV = 1 for a 1 mK edge.  Noise-only,
$\Lambda$-CDM-only maps were put through the PLANCK Edgefinder for
calibration purposes.  We made a total of 30 Stringy PLANCK maps with
only one string horizon per map.  In addition, 6 baseline
No-String maps were generated.  Since the range of EV for PLANCK is
expected to be wider, the optimum number of bins for the PLANCK data
is 704.  The total number of data points remain similar to the WMAP
simulation, so the comparison of the two statistics is valid.

\begin{figure}
\plotone{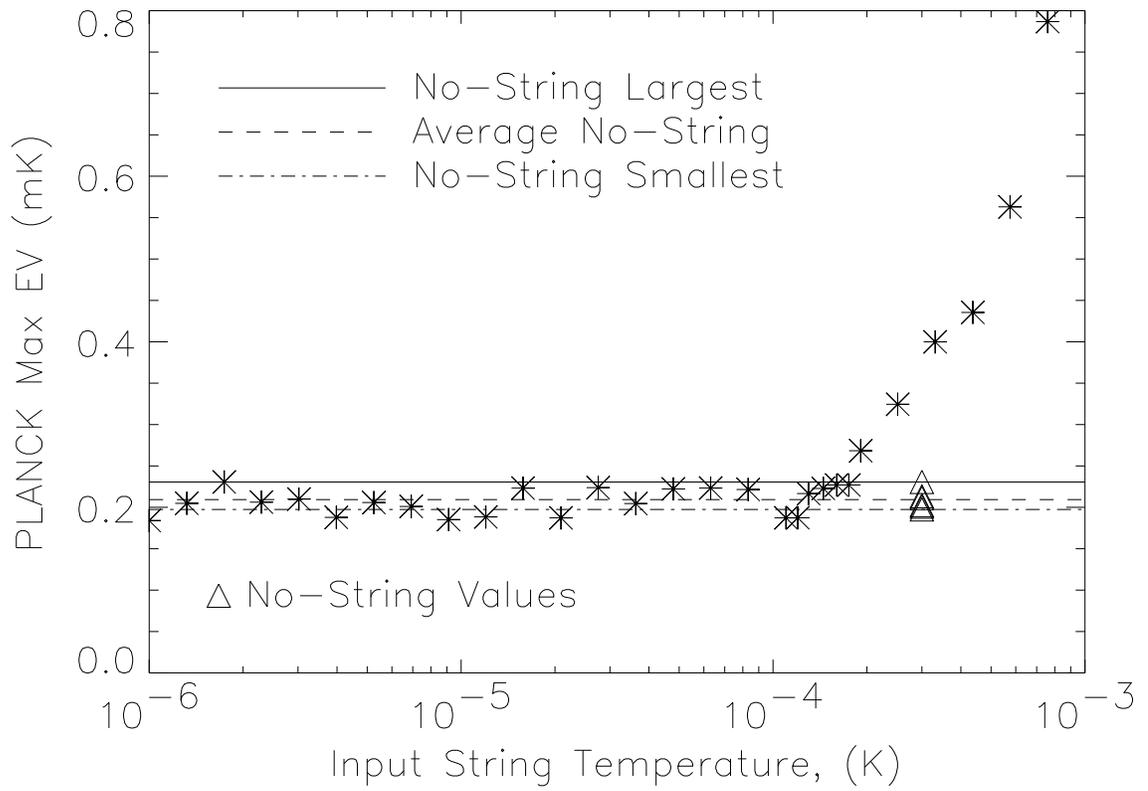}
\caption{Input string temperature vs. the maximum EV of the EV set for
simulated PLANCK data. \label{planck_max}}
\end{figure}

\begin{figure}
\plotone{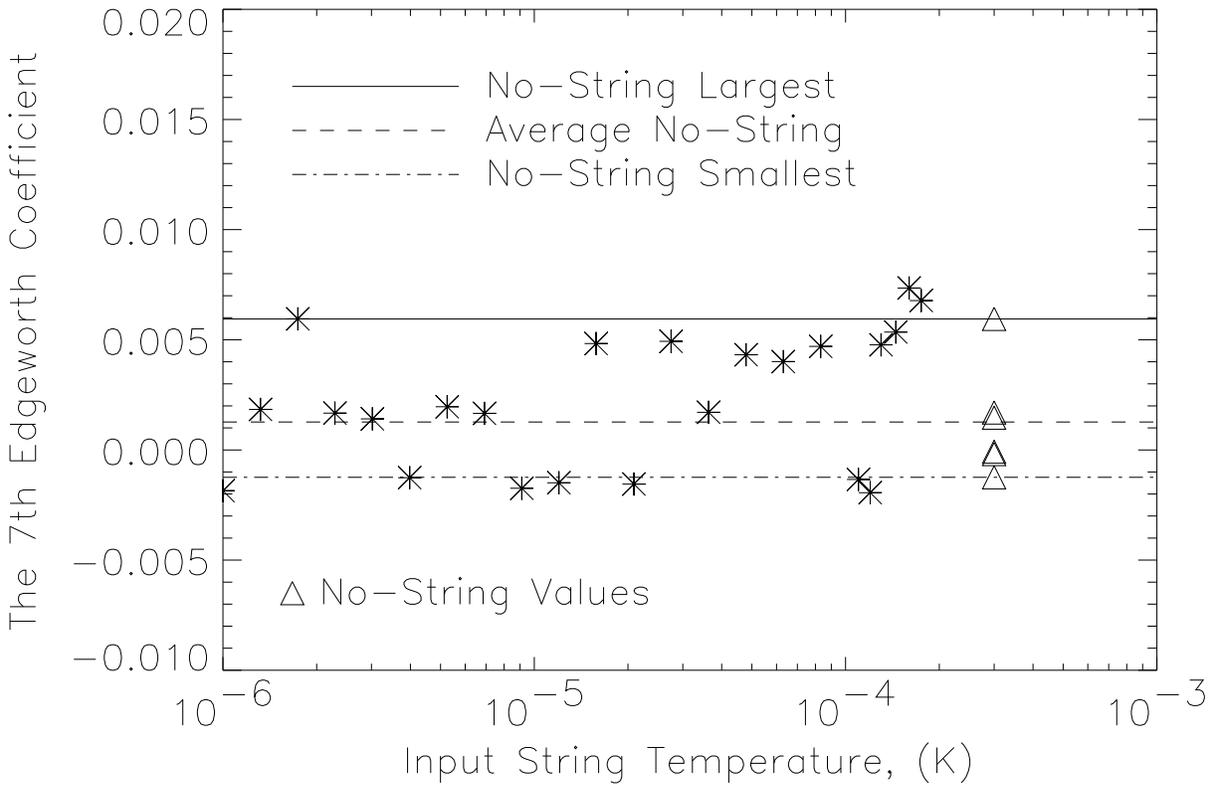}
\caption{Blow up of input string temperature vs. the 7th Edgeworth
coefficient of the EV set for simulated PLANCK data. \label{planck_e_7}}
\end{figure}

The plots of the string temperature, $T_s$, of a map vs. the max EV,
and vs. $E_7$ are in Figure \ref{planck_max} and Figure
\ref{planck_e_7}, respectively. The limit obtained from the max EV was
very similar to the the limit obtained from the $E_7$ coefficient,
most probably as a result of the fact that we have eliminated a lot of
the Gaussian signal by going to a smaller Edgefinder window.  This
strengthened our confidence in these limits. Both the Max EV and $E_7$
methods indicated that the detection limit for the PLANCK Edgefinder
occurred at T = 0.14 mK.  This represented a factor of 2 increase in
the single string Edgefinder sensitivity compared to the WMAP data.
This is, however, very similar to the multiple strings limit for the
WMAP data. For a cosmic string moving at mean simulation velocity,
$\beta = 1/\sqrt{2}$ this input string corresponds to a string mass of
$G\mu = 5.77 \times 10^{-6}$.

\section{Conclusions}

We have constructed and calibrated a digital filter for the WMAP CMB
data to search for cosmic strings.  By comparing the WMAP data and a
control set created from CMBFAST and SYNFAST, we have concluded that
the CMB data returned by the WMAP satellite do not contain single
strings to the limit of $G\mu \la 1.37 \times 10^{-5}$ using the max
EV as the threshold, and $G\mu < 1.07 \times 10^{-5}$ using the 7th
Edgeworth coefficient as the threshold.  This limit may be more
stringent if we allow the sky to have multiple strings, to a limit of
$G\mu < 5.97 \times 10^{-6}$, if there are more than 20 string
horizons in the sky.  We caution that WMAP effectively examined 2/3 of
the visible sky, so it is possible that we are missing strings in our
analysis. With the second year WMAP data, we can improve this limit by
about 5\%. This improvement is a result of an expected 30\%
improvement in the WMAP radiometer noise.  However, because the
Edgefinder filter window was 2 degrees, the first CMB doppler peak
generated a large background that cannot be filtered out, and
therefore a large reduction in radiometer noise generated a relatively
small improvement in the discrimination threshold.

We have also investigated claims of a possible cosmic string detection
of the object CSL-1, and found little evidence of a string at this
position.  For the proposed string mass of $G\mu = 3.86 \times
10^{-7}$, WMAP CMB temperatures at the location of CSL-1 would require
the string to have been moving very relativistically, with $v \sim
0.96c$.  We conclude that this is unlikely, and that much more
sensitive and higher angular resolution data would be needed for a
critical test of CSL-1.

We ran the Edgefinder through simulated PLANCK data.  As more
information about the PLANCK satellite becomes available, we can do a
more realistic modeling of the noise characteristics and therefore get
better limits on the strength of cosmic strings PLANCK can
detect. Currently, we have a projected limit of $G\mu \ga 5.77 \times
10^{-6}$, a factor of two better than the single string limit from
WMAP.  The predicted detector noise of PLANCK is too small to affect
the filter appreciably, and the true limit of the PLANCK data for the
Edgefinder is cosmic variance.  The factor of two improvement over the
first year WMAP data is mostly a result of the smaller filter window
which reduced signals from the first CMB doppler peak.

With more and more sensitive all sky CMB surveys, we can begin to set
firm experimental limits on the existence of GUT scale cosmic strings,
and thereby limiting the types of allowed phase transitions.

\acknowledgments
We would like to thank the authors of CMBFAST U. Seljak and
M. Zeldarriaga, and the creators of HEALPix (Gorski, Hivon, and
Wandelt 1999).

\begin{deluxetable}{cccc}
\tablecaption{Relevant WMAP Characteristics  \label{WMAP_char}}
\tablewidth{0pt}
\tablehead{
\colhead{Frequency} & \colhead{Resolution} & \colhead{Wavelength} &
\colhead{Sensitivity} \\
\colhead{GHz} & \colhead{FWHM, arcmin} & \colhead{mm} &
\colhead{$\mu$K, $0.3\degr \times 0.3\degr$} 
}
\startdata
23 & 56 & 13.6 & 35 \\ 
33 & 41 & 10.0 & 35 \\
41 & 32 &  7.5 & 35 \\
61 & 21 &  5.0 & 35 \\
94 & 13 &  3.3 & 35 \\
\enddata
\end{deluxetable} 

\begin{deluxetable}{ccc}
\tablecaption{Partially Exposed Strings  \label{exposure}}
\tablewidth{0pt}
\tablehead{
\colhead{Exposure \%} & \colhead{Normalized Max EV} &
\colhead{Non-zero Pixels} 
}
\startdata
100  & 0.961 & 575\\
73.6 & 0.956 & 339\\
48.7 & 0.569 & 210\\
32.6 & 0.160 & 135\\
18.0 & 0.143 & 71
\enddata
\end{deluxetable} 

\begin{deluxetable}{ccc}
\tablecaption{Edgefinder Gain Calibration \label{cali_val}}
\tablewidth{0pt}
\tablehead{
\colhead{$T_s$\tablenotemark{a}} & \colhead{EV Gain$|_{\alpha = 0}$\tablenotemark{b}} & 
\colhead{EV Gain$|_{\alpha = \pi/2}$}
}
\startdata
   1.000000  &   0.9567   &   0.9956 \\
   0.100000  &   0.9985   &   0.9732 \\
   0.010000  &   1.0800   &   0.9085 \\
   0.005000  &   0.8456   &   0.9630 \\
   0.001000  &   0.7639   &   1.0150 \\
   0.000500  &   0.8046   &   0.9842 \\
   0.000100  &   0.9296   &   0.8974 \\
   0.000050  &   0.8974   &   0.9794 \\
   0.000010  &   0.9559   &   1.0200 \\
   0.000001  &   1.0440   &   1.0910
\enddata
\tablenotetext{a}{$T_s$ is the input cosmic string horizon temperature}
\tablenotetext{b}{EV Gain is |EV/$T_s$|, here quoted for 2 different $\alpha$ angles.}
\end{deluxetable}

\begin{deluxetable}{ccccccc}
\tablecaption{Number and Strength of Input Strings for Multi-String
  Simulated Maps. \label{multi_s} }
\tablewidth{0pt}
\tablehead{
\colhead{Data Set} & \colhead{\# of Strings} & \colhead{$RAD$} &
\colhead{$T_s$} & \colhead{Max. EV} & \colhead{$E_7 (10^{-6})$}
}
\startdata
\bf{Multi A} &   2    &  2.0 &  0.30   & 0.290  &  $ 84.4  $ \\
\bf{Multi B} &   4    &  2.0 &  0.30   & 0.310  &  $ 185   $ \\
   Multi C   &   10   &  2.0 &  0.030  & 0.231  &  $ 2.86  $ \\
   Multi D   &   4    &  2.0 &  0.030  & 0.236  &  $-2.21  $ \\
   Multi E   &   30   &  2.0 &  0.030  & 0.257  &  $ 3.59  $ \\
   Multi F   &   60   &  2.0 &  0.030  & 0.243  &  $-2.42  $ \\
   Multi G   &   5    &  2.0 &  0.015  & 0.231  &  $ 2.09  $ \\
   Multi H   &   10   &  2.0 &  0.015  & 0.245  &  $ 4.76  $ \\
\bf{Multi I} &   20   &  2.0 &  0.15  & 0.244  &   $ 8.77  $ \\
   Multi J   &   2    &  4.0 &  0.015  & 0.249  &  $ 4.49  $ \\
   Multi K   &   2    &  4.0 &  0.030  & 0.263  &  $ 0.609 $ \\
\bf{Multi L} &   2    &  1.0 &  0.30   & 0.265  &  $ 46.2  $ \\
   Multi M   &   5    &  2.0 &  0.15   & 0.234  &  $ 0.394 $ \\
   Multi N   &   10   &  2.0 &  0.15   & 0.230  &  $ 2.54  $ \\
\bf{Multi O} &   5    &  1.0 &  0.30   & 0.294  &  $ 120   $ \\
\bf{Multi P} &   10   &  1.0 &  0.30   & 0.294  &  $ 114   $ \\
\enddata
\end{deluxetable}

\begin{deluxetable}{cccc}
\tablecaption{Select Properties of the WMAP composite QVW Data. \label{WMAP_stats} }
\tablewidth{0pt}
\tablehead{
\colhead{WMAP Band} & \colhead{Max Map T} & \colhead{Max EV} & \colhead{$E_7$}
}
\startdata
QVW  &     0.666   &   0.258   & $-3.82 \times 10^{-6}$ \\
\enddata
\end{deluxetable}

\begin{deluxetable}{cccc}
\tablecaption{Edgefinder Values for CSL-1 \label{tbl-1}}
\tablewidth{0pt}
\tablehead{
\colhead{Pixel \#} & \colhead{WMAP QVW T}   & \colhead{EV$|_{\alpha = 0}$}/\%-tile  &
\colhead{EV$|_{\alpha = \pi/20}$/\%-tile}
}
\startdata
968549  & -0.144015   &     0.07336 / 95.30\%  &  0.08686 / 97.61\% \\
&&& \\
968550  & -0.0735366  &     0.08322 / 97.11\%  &  0.08280 / 97.05\%\\
968548  &  0.0418802  &     0.1248  / 99.76\%   &  0.08621 / 97.53\%\\
968527  &  0.0717491  &     0.1234  / 99.74\%   &  0.09720 / 98.65\%\\
\enddata
\tablecomments{EV of CSL-1 and 3 immediately adjacent pixels.  Values listed
               are for $\alpha = 0$ and $\alpha = \pi/20$ }
\end{deluxetable}

\end{document}